\documentstyle[12pt]{article}
\newcommand{\spar}{{\hspace*{\parindent}}}
\newcommand{\sv}[1]{{\mbox{\boldmath $#1$}}}
\newfont{\msbm}{msbm10}
\newcommand{\RR}{\mbox{\msbm{\symbol{'122}}}}

\title{Prototype model for nuclear spin conversion in molecules: \\
The case of "hydrogen".}

\author
{{L.V.Il'ichov}
\thanks{
fax: (383-2)333-863}  \\
\normalsize{\it Institute of Automation and Electrometry SB RAS,}\\
\normalsize{\it Novosibirsk State University,}\\
\normalsize{\it Novosibirsk 630090, Russian Federation}}

\date{}

\begin{document}


\maketitle
\ \hrulefill \ \\


\begin{abstract}
        Using the conception of the so-called {\it quantum relaxation}, we build a semiphenomenological model for ortho-para conversion of nuclear spin isomers of hydrogen-type molecule. 
\end{abstract}    
\ \hrulefill \ \\ 
             


\section[Introduction]{Introduction}
\setcounter{equation}{0}
\spar
     
     
   Recently an idea was proposed [1] and proved [2-4] that the combined action of
intramolecular dynamics and environment could account for conversion of 
nuclear spin modifications of ${}^{13}CH_3F$. One may assume the same conversion
mechanism for other molecules. This mechanism resides in the following.
The intramolecular Hamiltonian is nondiagonal with respect to definite
total nuclear spin $I$. That implies the nutations between molecular states
with various $I$ values. The role of the environment is in the
braking of nutation's phase. Therefore the environment alone does not
initiate transitions with nuclear spin value changing, but, destructing
quantum coherence, the environment brakes the reversibility of nutations.
It is remarkable that no cross-section can be ascribed to such a conversion 
way, which makes pointing a classic analogy to this quantum process rather
problematic. For this reason, we name it {\it quantum
relaxation}. The known phenomena of enantiomers conversion [5],
decay of neutral kaons [6], and neutrino oscillations [7] are other examples of quantum relaxation.


   Considered as quantum relaxation, the nuclear spin conversion delivers 
a number of interesting problems, especially when being extended beyond the 
lowest approximations
with respect to the spin-mixing intramolecular Hamiltonian. This was shown
in the work [8], where quantum relaxation of multilevel system was considered
in the context of rather general approach. The price paid for the progress,
was also rather high. It consisted of specific form of intramolecular
Hamiltonian which mixed only a pare of states belonging to various spin
modifications. The dissipative term in the quantum evolution equation,
was particular as well, being a modification of BGK collision model.


   In the present work an attempt is made to overcome these limitations
in the special case of hydrogen-type molecule. The success of such an approach
is stipulated by specific algebraic properties of operators governing the
evolution of nuclear spin and molecular frame. The author states by no
means the quantum relaxation being the main (or even significant) cause
for conversion of real ortho- and para-hydrogen. We consider $H_2$-type-molecule 
because of the following three reasons:\\
1. due to its simplicity $H_2$-molecule turns to be an ideal testing site
for models of the influence of
the environment and for investigation of sensitivity of the conversion
rate to the variety of the models; \\
2. one can write the most general form for the intramolecular nuclear 
spin-mixing Hamiltonian allowable for $H_2$;\\
3. it is possible to account and to some extent investigate analitically higher orders
of the spin-mixing interaction in the conversion process.


   In Sec.2 and Sec.3 we introduce the set of operators for, respectively,
nuclear spin subsystem and molecular frame. In what follows (Sec.4 and below)
these operators are used as building blocks for construction of spin-mixing
Hamiltonian and various dissipative terms.
   


\section[Operator algebra of nuclear spin subsystem]{Operator algebra of
nuclear spin subsystem}
\setcounter{equation}{0}
\spar


   The operators acting in the nuclear spin space of $H_2$
molecule are expressed through the protons' spin operators
$\hat{\sv{I}}^{(1)}$ and $\hat{\sv{I}}^{(2)}$ and may be considered as
a basis of a 16-dimentional associative algebra. Let us introduce and describe
its elements. This algebra includes naturally the three
components of the total nuclear spin vector 
$\hat{\sv{I}} = \hat{\sv{I}}^{(1)} + \hat{\sv{I}}^{(2)}$. The total spin
magnitude may take two values - $0$ and $1$.
One may introduce the operator of the nuclear spin value $\hat{I}$:
\begin{equation}
\hat{I} = \frac{3}{4} + \left(\hat{\sv{I}}^{(1)}\cdot\hat{\sv{I}}^{(2)}\right).
\label{2.1}
\end{equation}
with the mentioned eigenvalues. This operator satisfies the evident
relation $\left(\hat{\sv{I}}\cdot\hat{\sv{I}}\right) = \hat{I}(\hat{I}+1)$.
We will either need raising and lowering operators with respect to $\hat{I}$.
Not too stretched manipulations provide the following vector operators
\begin{equation}
\hat{\sv{I}}^{(\pm)} = \hat{\sv{I}}^{(1)}-\hat{\sv{I}}^{(2)}
\pm 2i\left(\hat{\sv{I}}^{(1)}\times\hat{\sv{I}}^{(2)}\right)
\label{2.2}
\end{equation}
with the needed properties, which is evident from the commutators
\begin{equation}
[\hat{\sv{I}}^{(\pm)},\hat{I}] = \mp\hat{\sv{I}}^{(\pm)}.
\label{2.3}
\end{equation}
Repeated application of any rising and lowering operators' components 
gives zero:
\begin{equation}
\hat{I}^{(+)}_{i}\hat{I}^{(+)}_{j} =
\hat{I}^{(-)}_{i}\hat{I}^{(-)}_{j} = 0.
\label{2.4}
\end{equation}
One has also
\begin{equation}
\hat{I}\hat{\sv{I}}^{(-)} =
\hat{\sv{I}}^{(+)}\hat{I} = 0,
\label{2.5}
\end{equation}
and, respectively,
\begin{equation}
\hat{I}\hat{\sv{I}}^{(+)} = \hat{\sv{I}}^{(+)}, \;\;\;\;
\hat{\sv{I}}^{(-)}\hat{I} = \hat{\sv{I}}^{(-)}.
\label{2.6}
\end{equation}
The products of raising and lowering operators' components give
\[
\hat{I}^{(-)}_i\hat{I}^{(+)}_j = 4\delta_{ij}(1-\hat{I}),
\]
\begin{equation}
\label{2.7}
\end{equation}
\[
\hat{I}^{(+)}_i\hat{I}^{(-)}_j = \frac{4}{3}\delta_{ij}\hat{I} +
2i\epsilon_{ijk}\hat{I}_k - 4\hat{P}_{ij},
\]
where a new tensor operator
\begin{equation}
\hat{P}_{ij} = \hat{I}_{i}^{(1)}\hat{I}_{j}^{(2)} +
\hat{I}_{j}^{(1)}\hat{I}_{i}^{(2)} -
\frac{2}{3}\delta_{ij}\left(\hat{\sv{I}}^{(1)}\cdot\hat{\sv{I}}^{(2)}\right)
\label{2.8}
\end{equation}
appears. This operator is at the same time symmetric traceless tensor in
3D space. Supplemented by unit, the set 
$\{\hat{I}, \hat{\sv{I}}, \hat{\sv{I}}^{(+)}, \hat{\sv{I}}^{(-)}, \hat{P}_{ij}\}$
forms the associative algebra of the nuclear spin subsystem of $H_2$. 


   The following product relations involving $\hat{P}_{ij}$ 
will be needed:
$$
\hat{P}_{ij}\hat{I}^{(+)}_k = \frac{1}{3}\delta_{ij}\hat{I}^{(+)}_k -
\frac{1}{2}\delta_{jk}\hat{I}^{(+)}_i - \frac{1}{2}\delta_{ki}\hat{I}^{(+)}_j,
$$
$$
\hat{I}^{(-)}_k\hat{P}_{ij} = \frac{1}{3}\delta_{ij}\hat{I}^{(-)}_k -
\frac{1}{2}\delta_{jk}\hat{I}^{(-)}_i - \frac{1}{2}\delta_{ki}\hat{I}^{(-)}_j,
$$
\begin{equation}
\hat{P}_{ij}\hat{I}_{k}^{(-)} = \hat{I}_{k}^{(+)}\hat{P}_{ij} = 0.
\label{2.9}
\end{equation}
To finish the set of products which will be used, one should add
$$
\hat{I}_{i}\hat{I}_{j}^{(+)} = i\epsilon_{ijk}\hat{I}_{k}^{(+)}, ;\;\;
\hat{I}_{i}^{(-)}\hat{I}_{j} = i\epsilon_{ijk}\hat{I}_{k}^{(-)},
$$
\begin{equation}
\hat{I}_{i}\hat{I}_{j}^{(-)} = \hat{I}_{i}^{(+)}\hat{I}_{j} = 0.
\label{2.10}
\end{equation}
All other operator products are not of interest to us in the present
work.


\section[Operator algebra of molecular frame]{Operator algebra of
molecular frame}
\setcounter{equation}{0}
\spar


   It was shown in [9] that the evolution of molecular rotator
could be described within the framework of 10-dimentional space spanned 
by the operator set 
$\{\hat{J}, \hat{\sv{J}}, \hat{\sv{J}}^{(+)}, \hat{\sv{J}}^{(-)}\}$
which is at the same time the Lie algebra of the real simplectic group
$Sp(4,\RR)$ 
or, speaking more
precisely, of its simply connected covering metaplectic group $Mp(4,\RR)$
[10]. Here $\hat{\sv{J}}$ is the molecular rotational angular momentum;
$\hat{J}$ is the angular momentum operator value, so
that $(\hat{\sv{J}}\cdot \hat{\sv{J}}) = \hat{J}(\hat{J}+1)$;
$\hat{\sv{J}}^{(+)}$ and $\hat{\sv{J}}^{(-)} = \hat{\sv{J}}^{(+)\dagger}$
are vector operators
responsible, respectively, for transitions 
$J \mapsto J \pm 1$.


   The structure of the metaplectic Lie algebra is characterized by 
the following set of commutators:    
$$
[\hat{J}_{i},\hat{J}_{j}] = i\epsilon_{ijk}\hat{J}_{k},\;\; 
[\hat{J}_{i},\hat{J}^{(\pm)}_{j}] = i\epsilon_{ijk}\hat{J}^{(\pm)}_{k},\;\;
[\hat{J}_{i},\hat{J}] = 0; 
$$
\begin{equation}
[\hat{J}_{i}^{(+)},\hat{J}_{j}^{(-)}] =
-\delta_{ij}(2\hat{J}+1) - 2i\epsilon_{ijk}\hat{J}_k,
\label{3.1}
\end{equation}
$$
[\hat{J}_{i}^{(+)},\hat{J}_{j}^{(+)}] =
[\hat{J}_{i}^{(-)},\hat{J}_{j}^{(-)}] = 0, \;\;\;
[\hat{\sv{J}}^{(\pm)},\hat{J}] = \mp\hat{\sv{J}}^{(\pm)}.
$$
We need also the rules of $\hat{\sv{J}}^{(+)}$ and $\hat{\sv{J}}^{(-)}$
operation in the basis set of molecular rotational states
$\{|J,M\rangle\}$:
\begin{eqnarray}
\hat{J}_{+}^{(+)}|J,M\rangle& = &-\sqrt{(J+M+1)(J+M+2)}|J+1,M+1\rangle,\nonumber \\
\hat{J}_{-}^{(+)}|J,M\rangle& = &\;\sqrt{(J-M+1)(J-M+2)}|J+1,M-1\rangle, \nonumber \\
\hat{J}_{3}^{(+)}|J,M\rangle& = &\;\sqrt{(J+M+1)(J-M+1)}|J+1,M\rangle, \nonumber\\
\label{A.5} \\
\hat{J}_{+}^{(-)}|J,M\rangle& = &\;\sqrt{(J-M)(J-M-1)}|J-1,M+1\rangle, \nonumber\\
\hat{J}_{-}^{(-)}|J,M\rangle& = &-\sqrt{(J+M)(J+M-1)}|J-1,M-1\rangle, \nonumber\\
\hat{J}_{3}^{(-)}|J,M\rangle& = &\;\sqrt{(J+M)(J-M)}|J-1,M\rangle, \nonumber
\end{eqnarray}
where $\hat{J}^{(+)}_{\pm} = \hat{J}^{(+)}_1 \pm i\hat{J}^{(+)}_2$,
$\hat{J}^{(-)}_{\pm} = \hat{J}^{(-)}_1 \pm i\hat{J}^{(-)}_2$. From these
relations one can easily prove the validity of the following relations 
which will be widely used below:
$$
(\hat{\sv{J}}\times\hat{\sv{J}}^{(+)}) = i(\hat{J}+1)\hat{\sv{J}}^{(+)},\;\;
(\hat{\sv{J}}^{(+)}\times\hat{\sv{J}}) = - i\hat{\sv{J}}^{(+)}\hat{J},
$$
$$
(\hat{\sv{J}}\times\hat{\sv{J}}^{(-)}) = -i\hat{J}\hat{\sv{J}}^{(-)},\;\;
(\hat{\sv{J}}^{(-)}\times\hat{\sv{J}}) = i\hat{\sv{J}}^{(-)}(\hat{J}+1).
$$
$$
(\hat{\sv{J}}^{(+)} \cdot \hat{\sv{J}}^{(-)}) = \hat{J}(2\hat{J}-1), \;\;
(\hat{\sv{J}}^{(-)} \cdot \hat{\sv{J}}^{(+)}) = (\hat{J}+1)(2\hat{J}+3);
$$
\begin{equation}
(\hat{\sv{J}}^{(+)}\times\hat{\sv{J}}^{(-)}) = i(2\hat{J}-1)\hat{\sv{J}},\;\; 
(\hat{\sv{J}}^{(-)}\times\hat{\sv{J}}^{(+)}) = -i(2\hat{J}+3)\hat{\sv{J}}; 
\label{3.2}
\end{equation}
$$
\hat{J}_i^{(+)}\hat{J}_j^{(-)} = \hat{J}^2\delta_{ij} + 
i\hat{J}\epsilon_{ijk}\hat{J}_k - \hat{J}_i\hat{J}_j;
$$
$$
\hat{J}_i^{(-)}\hat{J}_j^{(+)} = (\hat{J}+1)^2\delta_{ij} - 
i(\hat{J}+1)\epsilon_{ijk}\hat{J}_k - \hat{J}_i\hat{J}_j;
$$
$$
(\hat{\sv{J}}^{(+)}\cdot \hat{\sv{J}}^{(+)}) =
(\hat{\sv{J}}^{(-)}\cdot \hat{\sv{J}}^{(-)}) =
(\hat{\sv{J}}\cdot \hat{\sv{J}}^{(+)})=
(\hat{\sv{J}}\cdot \hat{\sv{J}}^{(-)}) = 0
$$
The last line in (\ref{3.2}) has very clear physical meaning. It states that
no scalar operator, which acts only in the space of molecular rotation states,
can generate transitions with $\Delta J \neq 0$. 


\section[Mixing intramolecular interaction]{Mixing intramolecular
interaction}
\setcounter{equation}{0}
\spar


   For simplicity the model molecular dynamics we are going to consider
consists of free molecular rotation with a nuclear spin-mixing interaction
imposed on. Vibrational and electron degrees of freedom are not of interest 
and the vibrational state is assumed to be symmetric. 


   Let us specify the form of the spin-mixing Hamiltonian $\hat{H}_{mix}$.
Because of Pauli principle, ortho-hydrogen $(I=1)$ can have only odd
rotational rotational angular momentum values $J$, whereas para-hydrogen
$(I=0)$ must be in a rotational state with even $J$. There must be operators
in $\hat{H}_{mix}$, which generate transitions between ortho- and para-states.
$\hat{\sv{I}}^{(\pm)}$ from Sec.2 are the only operators of such a kind.
Because of (\ref{2.4}), the components of these operators may be
contained in $\hat{H}_{mix}$ in the first degree only. The mentioned 
correlation between $I$ and $J$ should not be violated by $\hat{H}_{mix}$.
This makes $\hat{\sv{I}}^{(\pm)}$ be multiplied by vector operators
affecting the molecular rotation by changing $J$ by an odd number. We
note that any vector operator, which changes $J$ by 3 or more, must
inevitably contain the scalar products 
$(\hat{\sv{J}}^{(+)}\cdot\hat{\sv{J}}^{(+)})$ or      
$(\hat{\sv{J}}^{(-)}\cdot\hat{\sv{J}}^{(-)})$ and should be equal to zero
in accordance with (\ref{3.2}). Hence $\hat{\sv{J}}^{(\pm)}$ are the only
operators which $\hat{\sv{I}}^{(\pm)}$ may be multiplied by. One can now
write the general form of the mixing Hamiltonian:
\begin{eqnarray}
\hat{H}_{mix} = &\omega_{+}(\hat{J})(\hat{\sv{J}}^{(+)}\cdot\hat{\sv{I}}^{(+)})\;\;
+&(\hat{\sv{J}}^{(-)}\cdot\hat{\sv{I}}^{(-)})\bar{\omega}_{+}(\hat{J})\nonumber \\
+&\omega_{-}(\hat{J})(\hat{\sv{J}}^{(+)}\cdot\hat{\sv{I}}^{(-)})\;\;
+&(\hat{\sv{J}}^{(-)}\cdot\hat{\sv{I}}^{(+)})\bar{\omega}_{-}(\hat{J}).
\label{4.1}
\end{eqnarray}
The scale factors $\omega_{\pm}(\hat{J})$ in (\ref{4.1}) account for $J$-
dependence of nuclear spin-mixing and can be specified by a microscopic
model of mixing interaction. The line over the symbols stands for complex
conjugation. Taken with the rotational Hamiltonian
$\hat{H}_0 = \omega_0\hat{J}(\hat{J}+1)$, the terms (\ref{4.1}) add up
to the total molecular Hamiltonian $\hat{H}$ which will be used below.


\section[Heisenberg operator for nuclear spin value $\hat{I}(t)$]
{Heisenberg operator for nuclear spin value $\hat{I}(t)$}
\setcounter{equation}{0}
\spar


   Written in the superoperator form, the kinetic equation for the 
molecules' density matrix reads
\begin{equation}
\partial_t\hat{\rho}(t) = {\cal L}[\hat{\rho}(t)] \equiv 
{\cal L}_0[\hat{\rho}(t)] + \sum_{n=1}^{N}\nu_n{\cal L}_n[\hat{\rho}(t)],
\label{5.1}
\end{equation}   
where two types of superoperators (Liouvillians) occur:
${\cal L}_0[\hat{\rho}(t)] \equiv -i[\hat{H}, \hat{\rho}(t)]$ is the
Liouvillian of the free Hamiltonian dynamics; ${\cal L}_n$ are the 
generators of "irreversible parts" of evolution caused by the environment.
Various dissipative generators contribute additively and are specified
by the number $n = 1,2,\ldots,N$; $\nu_n$ give the rates of corresponding
dissipative processes. In the present work we are going to describe
three types of dissipative Liouvillians and consider the simplest two
ones more or less comprehensively.


   In the context of the simplest approach, one assumes that any collision
with the environment's particles projects the hydrogen molecule onto
pure ortho- and para-states; and this is the only result of the collision
in the considered model. Hence
\begin{equation}
{\cal L}_1[\hat{\rho}] = \hat{I}\hat{\rho}\hat{I} +
(1-\hat{I})\hat{\rho}(1-\hat{I}) -\hat{\rho} \equiv
[\hat{I},[\hat{\rho},\hat{I}]]
\label{5.2}
\end{equation}   
One can easily prove that ${\cal L}_1$ causes the extinction of 
ortho-para-states coherence induced by the mixing Hamiltonian. 


   In contrast to the first model, the second one is formulated in
terms of molecular frame operators. It postulates that collisions with the 
environment's particles cause frequent infinitesimal rotations of $\sv{J}$ 
(but do not affect directly the nuclear spin):
\begin{equation}
{\cal L}_2[\hat{\rho}] \equiv \hat{\sv{J}}\hat{\rho}\hat{\sv{J}} -
\frac{1}{2}\hat{J}(\hat{J}+1)\hat{\rho} -
\frac{1}{2}\hat{\rho}\hat{J}(\hat{J}+1)
\label{5.3}
\end{equation}
Note that the ortho-para-states coherence 
means at the same time the coherence between states with various
$J$ (due to the mentioned $I-J$ correlation). Because of the last
two terms in (\ref{5.3}) the deorientational collisions destruct 
the coherence. 


   The third dissipative model accounts for collisional transitions 
changing $J$. In every collision one has $J \mapsto J\pm2$. So
the collisions do not initiate direct transitions between ortho- and
para-states. The corresponding Liouvillian has the form
\begin{eqnarray}
{\cal L}_3[\hat{\rho}]& = &[\hat{J}_{ij}^{(-)}\bar{q}(\hat{J})\exp\{\beta(2\hat{J}-1)\}\hat{\rho},
q(\hat{J})\hat{J}_{ij}^{(+)}]  \\ \nonumber
& + &[q(\hat{J})\exp\{-\beta(2\hat{J}-1)\}\hat{J}_{ij}^{(+)}\hat{\rho},
\hat{J}_{ij}^{(-)}\bar{q}(\hat{J})] + H.c.,
\label{5.100}
\end{eqnarray}
where $\hat{J}_{ij}^{(\pm)} \equiv \hat{J}_i^{(\pm)}\hat{J}_j^{(\pm)}$ are
traceless symmetric tensors; $\beta = \hbar\omega_0/k_BT$; $q(\hat{J})$ is a
function of $J$. This model is a strait analog of that one from the work
[9]. The conversion induced by (\ref{5.100}) will be considered in
details elsewhere.


   Arming with ${\cal L}_1$ and ${\cal L}_2$, we may return to Eq.(\ref{5.1}).
It is more convenient to deal with Heisenberg observable $\hat{O}(t)$
rather than with $\hat{\rho}(t)$. The corresponding equation of motion
for $\hat{O}(t)$ reads
\begin{equation}
\partial_t\hat{O}(t) = {\cal L}^{\dagger}[\hat{O}(t)],
\label{5.4}
\end{equation}    
where the adjoint Liouvillian is introduced. It is determined with respect
to the trace scalar product 
$Tr(\hat{O}{\cal L}[\hat{\rho}]) = Tr({\cal L}^{\dagger}[\hat{O}]\hat{\rho})$.
Note that ${\cal L}_0^{\dagger} = -{\cal L}_0$, ${\cal L}_1^{\dagger} =
{\cal L}_1$, and ${\cal L}_2^{\dagger} = {\cal L}_2$.


   Of concern to us is the Heisenberg operator $\hat{I}(t)$ of the
nuclear spin value. The detailed analysis shows that $\hat{I}(t)$ (as
well as any scalar Heisenberg operator) has the following structure:
$$
\hat{I}(t) = A(\hat{J},t) + B(\hat{J},t)\hat{I} +
C^{(+)}(\hat{J},t)(\hat{\sv{J}}^{(+)}\cdot\hat{\sv{I}}^{(+)})
$$
\begin{equation}
+ C^{(-)}(\hat{J},t)(\hat{\sv{J}}^{(+)}\cdot\hat{\sv{I}}^{(-)})
+ (\hat{\sv{J}}^{(-)}\cdot\hat{\sv{I}}^{(-)})\bar{C}^{(+)}(\hat{J},t)
+ (\hat{\sv{J}}^{(-)}\cdot\hat{\sv{I}}^{(+)})\bar{C}^{(-)}(\hat{J},t)
\label{5.5}
\end{equation}    
$$
+ D(\hat{J},t)(\hat{\sv{J}}\cdot\hat{\sv{S}}) 
+ E(\hat{J},t)\hat{J}_i\hat{P}_{ij}\hat{J}_j
+ F(\hat{J},t)\hat{J}_i^{(+)}\hat{P}_{ij}\hat{J}_j^{(+)}
+ \hat{J}_i^{(-)}\hat{P}_{ij}\hat{J}_j^{(-)}\bar{F}(\hat{J},t).
$$
The coefficients $A(\hat{J},t)$, \ldots, $F(\hat{J},t)$ are to be
determined.   


   With the formal solution of Eq.(\ref{5.4}) presented as the
Taylor expansion
\begin{equation}
\hat{O}(t) = \sum_{n=0}^{\infty}\frac{t^n}{n!}({\cal L}^{\dagger})^n[\hat{O}],   
\label{5.6}   
\end{equation}
one may associate the expansions
\begin{equation}
A(\hat{J},t) = \sum_{n=0}^{\infty}\frac{t^n}{n!}A_n(\hat{J}), \ldots,
F(\hat{J},t) = \sum_{n=0}^{\infty}\frac{t^n}{n!}F_n(\hat{J}).
\label{5.7}
\end{equation}
After rather cumbersome calculations with the use of algebraic relations
from Secs.2 and 3, we arrive at the following set of itterative equations:
\begin{equation} 
A_{n+1}(J) = -4(J+1)(2J+3)L_n^{(+)}(J+1) + 4J(2J-1)L_n^{(-)}(J);
\label{5.8}
\end{equation}\\
\begin{equation} 
B_{n+1}(J) = \frac{4}{3}J(2J-1)[L_n^{(+)}(J) - 3L_n^{(-)}(J)]
- \frac{4}{3}(J+1)(2J+3)[L_n^{(-)}(J+1) - 3L_n^{(+)}(J+1)];
\label{5.9}
\end{equation}  \\
\begin{equation}
C_{n+1}^{(+)}(J) = - \frac{i}{6}(J+1)(2J+3)\omega_{+}(J)E_n(J)
+ i(J-1)(2J-3)\bar{\omega}_{-}(J-1)F_n(J)
\label{5.10}
\end{equation}
$$
+i\omega_{+}(J)[A_n(J-1) - A_n(J)] - i\omega_{+}(J)B_n(J)
+ [2i\omega_0J - \nu_1 - \nu_2]C_n^{(+)}(J) + i\omega_{+}(J)(J+1)D_n(J);
$$\\ 
\begin{equation}
C_{n+1}^{(-)}(J) = \frac{i}{6}(J-1)(2J-3)\omega_{-}(J)E_n(J-1)
- i(J+1)(2J+3)\bar{\omega}_{+}(J+1)F_n(J+1)
\label{5.11}
\end{equation}
$$
- i\omega_{-}(J)[A_n(J-1) - A_n(J)] + i\omega_{-}(J)B_n(J) + [2i\omega_0J - \nu_1 - \nu_2]C_n^{(-)}(J) + i\omega_{-}(J)(J-1)D_n(J);
$$\\
\begin{equation}
D_{n+1}(J) = -2(2J-1)L_n^{(+)}(J) -2(2J+3)L_n^{(-)}(J+1) - \nu_2D_n(J);
\label{5.12}
\end{equation}\\
\begin{equation}
E_{n+1}(J) = 4L_n^{(+)}(J) - 4L_n^{(-)}(J+1) - 3\nu_2E_n(J);
\label{5.12}
\end{equation}\\
\begin{equation}
F_{n+1}(J) = [2i(2J-1)\omega_0 - 3\nu_2]F_n(J)  
+ 4i\omega_{-}(J-1)C_{n}^{(+)}(J) - 4i\omega_{+}(J)C_n^{(-)}(J-1),
\label{5.13}
\end{equation}\\
where the combinations
$$
L_n^{(+)}(J) = -i\bar{\omega}_{+}(J)C_n^{(+)}(J) +
i\omega_{+}(J)\bar{C}_n^{(+)}(J)
$$
and
$$
L_n^{(-)}(J) = -i\bar{\omega}_{-}(J)C_n^{(-)}(J) +
i\omega_{-}(J)\bar{C}_n^{(-)}(J)
$$
are introduced. In the case of $\hat{I}(t)$ we must use the initial
conditions when all coefficients $A_0(J),\;\ldots$, $F_0(J)$ are zero
except $B_0(J) = 1$. In principle, Eqs.(\ref{5.8}-\ref{5.13}) let 
one evaluate $\hat{I}(t)$
provided the Taylor expansions converge.


\section[Conclusion]
{Conclusion}
\setcounter{equation}{0}
\spar


	Fundamentally, the Heisenberg operator $\hat{I}(t)$ let one evaluate any nuclear-spin-dependent magnitude. For example, the expression
\begin{equation}
\langle I(t)\rangle =  Tr\hat{I}(t)\hat{\rho}
\label{6.1}
\end{equation}
gives the conversion process in an initially prepared non-equilibrium state, where $\hat{\rho}$ is the initial density matrix. One  can easily show that (due to the I-J correlations) any relevant $\hat{\rho}$ must obey the 
equality
\begin{equation}
\left(2\hat{I} + (-1)^{\hat{J}}\right)\hat{\rho} = \hat{\rho} =
\hat{\rho}\left(2\hat{I} + (-1)^{\hat{J}}\right)
\label{6.2}
\end{equation}   
The stationary density matrix annulates both the left- and right-hand sides of (\ref{5.1}). It has a
form akin to (\ref{5.5}) with $t$-independent coefficients. The
simplest acceptable (but not stationary) isotropic density matrix is
\begin{equation}
\hat{\rho} = \rho_0(\hat{J}) + \rho_1(\hat{J})\hat{I}
\label{5.15}
\end{equation}
where in accordance with (\ref{6.2}) $\rho_0(J)=0$ for odd $J$ and
$\rho_0(J) = - \rho_1(J)$ for even $J$.


	The proposed model allows rather extended analitical evaluations. Numerical calculations are still inevitable. For this one should specify the factors $\omega_{\pm}(\hat{J})$ in (\ref{4.1}). This is an independent problem. There can be faced a situation of bad convergency of the Taylor expansions (\ref{5.8}) for some $\omega_{\pm}(\hat{J})$. The analysis of these and related problems will be done elsewhere. As a result we will be able to estimate the value of the model completely.\\[1cm]


{\bf Acknowledgements}


	This work was carried out while the author's stay in the Huygens Laboratory of Leiden University. This visit was funded by the Netherlands Organization for Scientific Research. The author gratefully acknowledges support and hospitality of Prof. L.J.F.Hermans.
	The author is also deeply indebted to P.L.Chapovsky for valuable remarks and discussions. Partial support from the Russian Foundation for Basic Research (grant N 98-03-33124a) and Federal Program "Integration" (project N 274) is acknowledged.


\end{document}